\documentclass[12pt]{article}
\usepackage{array}
\usepackage{graphicx}
\usepackage{amssymb}
\usepackage{amsmath}
\input{epsf}
\usepackage{cite}
\def\@fmsl@sh#1#2#3{\m@th\ooalign{$\hfil#1\mkern#2/\hfil$\crcr$#1#3$}}
 \def\eq#1\en{\begin{equation}#1\end{equation}}
\def\s[#1,#2]{[#1\stackrel{\star}{,}#2]}
\def\sx[#1,#2]{[#1\stackrel{\star_{x}}{,}#2]}

\newcommand{\nc}{\newcommand}
\nc{\beq}{\begin{equation}}
\nc{\eeq}{\end{equation}}
\nc{\beqa}{\begin{eqnarray}}
\nc{\eeqa}{\end{eqnarray}}

\def\bc{\begin{center}}
\def\ec{\end{center}}
\def\ie{{\it i.e.}}
\def\eg{{\it e.g.}}

\def\to{\rightarrow}

\def\gsim{\mathrel{\mathpalette\atversim>}}

\def\bc{\begin{center}}
\def\ec{\end{center}}

\def\gsim{\mathrel{\rlap{\lower4pt\hbox{\hskip1pt$\sim$}}

    \raise1pt\hbox{$>$}}}       %greater than or approx. symbol

\def\gsim{\mathrel{\rlap{\lower4pt\hbox{\hskip1pt$\sim$}}
    \raise1pt\hbox{$>$}}}       %greater than or approx. symbol

\begin{document}
\makeatletter
\def\fmslash{\@ifnextchar[{\fmsl@sh}{\fmsl@sh[0mu]}}
\def\fmsl@sh[#1]#2{%
  \mathchoice
    {\@fmsl@sh\displaystyle{#1}{#2}}%
    {\@fmsl@sh\textstyle{#1}{#2}}%
    {\@fmsl@sh\scriptstyle{#1}{#2}}%
    {\@fmsl@sh\scriptscriptstyle{#1}{#2}}}
\def\@fmsl@sh#1#2#3{\m@th\ooalign{$\hfil#1\mkern#2/\hfil$\crcr$#1#3$}}
\makeatother
%\baselineskip 24pt

%%%%%%%%%%%%%%%%%%%%%%%%%%%%%%%%%%%%%%%%%%%%%%%%%%%%%%%%%%%%%%%%%
%%%
%%%                      TITLE PAGE
%%%
%%%%%%%%%%%%%%%%%%%%%%%%%%%%%%%%%%%%%%%%%%%%%%%%%%%%%%%%%%%%%%%%%
\thispagestyle{empty}
\begin{titlepage}
\begin{flushright}
SLAC-PUB-13771\\
CP3-09-45
\end{flushright}
\vspace{0.3cm}
\boldmath
\begin{center}
  \Large {\bf Massless versus Kaluza-Klein Gravitons at the LHC}
\end{center}
\unboldmath
\vspace{0.8cm}
\begin{center}
{{\large Xavier Calmet}\footnote{x.calmet@sussex.ac.uk}$^{a}$, 
{\large Priscila de Aquino}\footnote{priscila@itf.fys.kuleuven.be}$^{b,c}$ \large{and Thomas G. Rizzo} \footnote{rizzo@slac.stanford.edu}$^d$
}
 \end{center}
\begin{center}
{\sl $^a$ Physics and Astronomy, 
University of Sussex,  \\ Falmer, Brighton, BN1 9QH, UK \\
 $^b$ Universit\' e Catholique de Louvain, CP3, \\
2, Chemin du Cyclotron,
B-1348 Louvain-la-Neuve, Belgium\\ 
$^c$ Katholieke Universiteit Leuven, ITP, \\
Celestijnenlaan 200D - Bus 2415,
B-3001 Leuven, Belgium \\
$^d$ SLAC National Accelerator Laboratory, \\
2575 Sand Hill Rd., 
Menlo Park, CA, 94025, USA 
}
\end{center}
\vspace{\fill}
\begin{abstract}
\noindent
We show that the LHC will be able to differentiate between a four-dimensional model with quantum gravity at $\sim 1$ TeV where the (massless) graviton becomes strongly coupled to standard model particles at 1 TeV and brane world type models with a large extra-dimensional volume and massive Kaluza-Klein gravitons. We estimate that the 14 TeV LHC could put a limit of the order of $\sim 5$ TeV on the four dimensional Planck mass in a model independent way.
\end{abstract}  
\end{titlepage}

%\pacs{}

%%%%%%%%%%%%%%%%%%%%%%%%%%%%%%%%%%%%%%%%%%%%%%%%%%%%%%%%%%%%%%%%
%%%
%%%                     INTRODUCTION
%%%
%%%%%%%%%%%%%%%%%%%%%%%%%%%%%%%%%%%%%%%%%%%%%%%%%%%%%%%%%%%%%%%%

\newpage

It is now well understood that the energy scale at which quantum gravitational effects become strong could be much lower than naively estimated 
using dimensional analysis arguments. For example in models with a brane world and a large extra dimensional 
volume  \cite{ArkaniHamed:1998rs, Randall:1999ee} or in four dimensional models with a large hidden sector of some $10^{33}$ 
particles \cite{Calmet:2008tn}\footnote{See also\cite{Dvali} for another discussion of the value of the Planck mass in models with a large hidden sector. In their approach it is not clear how the value of the Planck mass depends on the spin of the particles in the hidden sector. However, reliable perturbative calculations \cite{Larsen:1995ax} clearly show a spin dependence: spin 0 and 1/2  particles lower the Planck mass, while spin 1 particles increase it.} , the Planck mass could be near 1 TeV and thus in reach of colliders such as the Tevatron or the LHC.
In this paper we compare the phenomenology of the extra-dimensional brane world models (\ie, ADD type models) with that of the four-dimensional one. In particular we show that the LHC can differentiate between the two scenarios. Here we follow the strategy developed in \cite{Rizzo:2008vr} 
to differentiate the phenomenology of Kaluza-Klein gravitons in ADD models from that of unparticles \cite{georgi,Cheung:2008xu}.

The specific four-dimensional model discussed in \cite{Calmet:2008tn} is rather different at the theoretical level from the well known ADD model. Gravity can become strong in four dimensions if there is a large hidden sector that only interacts gravitationally with the standard model. 
This large hidden sector will contribute to the renormalization group running of Newton's constant. If the new particles in this sector 
have spin 0 and/or 1/2 the running of Newton's constant is such that it becomes significantly larger at lower energy. For $10^{33}$ new particles one gets $M_P(1$ TeV) $\sim 1$ TeV. The masses of these new particles determine the energy scale at which the running of Newton's constant starts, for example one could choose $m_i \sim 100$ GeV. In such a case most of the running of the Planck mass (equivalently of Newton's constant) occurs close to 1 TeV. One can thus approximate the running of the reduced Planck mass by a step function: for $\mu < 1 $TeV, $M(\mu)=2.4 \times 10^{18}$ GeV, while for $\mu \ge 1 $TeV, $M(\mu) \sim 1$ TeV. The phenomenology of this model is then found to be rather different from that of ADD since strong gravity cuts off at the TeV scale. For example, there are no Kaluza-Klein excitations of the graviton and virtual gravitons will decay massively to the hidden sector and are thus leading to a missing energy signal. 
It has been suggested in  \cite{Calmet:2009gn} to search for a monojet plus missing energy as a signature 
of this model at the LHC. This is, however, also one of the familiar smoking gun signals of the ADD scenario and thus it is important to establish that the LHC could differentiate between the two models. Furthermore, in case of non-observation of the monojet + missing energy signals, one could use data from the LHC to place a corresponding lower bound on the value of the Planck mass (i.e. the scale at which quantum gravitational effects become important). It is the aim of the present paper to establish these two points.{\footnote {Also, unlike in the ADD case, the exchange of these massless gravitons does not lead to effective dim-8 contact interactions with an associated characteristic TeV scale. Furthermore, there will be no observable alteration in Newtonian Gravity observable in table top experiments as may be possible with ADD in the case of 2 extra dimensions.}}

The production of jets with large $E_T$ recoiling against a {\it massless} graviton $G$ can arise from the  parton subprocesses  $q+ \bar q \to G + g$, $q+ g \to q+ G$, $\bar q + g \to \bar q + G$, and $g+g \to g+ G$ in a manner similar to that in ADD. However, the expressions for the corresponding subprocess cross-sections are different from those found in the ADD model. 

The leading order contributions at the parton level for this four dimensional model have been calculated and presented in \cite{Calmet:2009gn}. 
 The polarization and color averaged cross section for $q + \bar q \to g +G$ is given by
\begin{eqnarray}
 \frac{d\sigma}{d \cos\theta}&=& \frac{1}{72 \pi}
 \frac{g_s^2 }{\bar M(\mu)^2} \left ( \frac{s^2 +2t^2+ 2 t s}{s^2} \right ),
 \end{eqnarray}
where $g_s^2=4\pi \alpha_s$ is the usual strong coupling constant, hereafter evaluated at the scale of the jet $E_T$, 
$\bar M(\mu)$ is the reduced Planck mass such that Newton's constant is given by $G_N=(4\pi {\bar M}^2)^{-1}$ and where $s$ and $t$ are the subprocess Mandelstam variables: $t= -1/2 s (1 - \cos \theta)$.
The  cross sections for $q + g \to q + G$ and $\bar q + g \to \bar q + G$  are given by
 \begin{eqnarray}
\frac{d \sigma}{d \cos \theta} = - \frac{g_s^2 (2 s^2+ 2 s t + t^2)}{ 192 \pi \bar M(\mu)^2 s t }.
\end{eqnarray}

The corresponding matrix element can be obtained using crossing symmetry from that of the transition $q +\bar q \to G + g$.
Finally we also obtain the cross section for  $g+ g \to g + G$ which is given by
 \begin{eqnarray}
\frac{d \sigma}{d \cos \theta} = - \frac{3 g_s^2 (s^2+s t + t^2)^2}{ 128 \pi \bar M(\mu)^2 s^2 t (s+t)}.
\end{eqnarray}
Note that all quarks are treated as massless in these expressions. Recall that these cross sections are essentially zero below the $\sqrt s=M_P$ 
threshold. 
 In the case of ADD, the parton level cross-sections have been given in  \cite{Mirabelli:1998rt} and we quote them here for comparison purposes 
for KK gravitons of mass $m$.  The cross section $q + \bar q \to g + G_{KK}$ where $G_{KK}$ is an ADD Kaluza-Klein graviton is given by: 
\begin{eqnarray}
\frac{d\sigma}{d \cos\theta}(q + \bar q \to g + G_{KK})  &=& \frac{1}{144 \pi}
 \frac{g_s^2 }{\bar M^2}\frac{1}{1-m^2/s}\Biggl[(2 - \frac{4ut}{(s-m^2)^2}) 
  \left(1 + \bigl(\frac{m^2}{s}\bigr)^4\right) \\ \nonumber
     &&  + \left(2 \frac{ (s-m^2)^2}{4 ut} - 5 
      + 4 \frac{4ut}{(s-m^2)^2}\right)\frac{m^2}{s } \left(1 +  
   \bigl(\frac{m^2}{s}\bigr)^2\right) + 
   \\ \nonumber
     && + 6 \left( \frac{u-t}{s-m^2}\right)^2
                          \bigl(\frac{m^2}{s}\bigr)^2
                                          \Biggr] \ ,
\end{eqnarray}
where here $s,t,u$ are the Mandelstam variables where we have: $t,u = -1/2 s (1-m^2/s)(1\mp \cos \theta)$.
The cross section for $q+g \to q+ G_{KK}$  can be obtained from this expression by crossing $s \leftrightarrow t$:
\begin{eqnarray}
\frac{d\sigma}{d \cos\theta}(q+g \to q+ G_{KK})  &=&  
\frac {g_s^2}{384 \pi \bar M^2} \frac{(-t/s)(1-m^2/s)}{(1-m^2/t)^2} \times \\
\nonumber
     && 
 \times \Biggl[(2 - \frac{4us}{ (t-m^2)^2}) 
  \left(1 + \bigl(\frac{m^2}{t}\bigr)^4\right) \\ \nonumber
     && + \left(2 \frac{ (t-m^2)^2}{4 us} - 5 
      + 4 \frac{4us}{(t-m^2)^2}\right)\frac{m^2}{t }\left(1 +  
   \bigl(\frac{m^2}{t}\bigr)^2\right) +\\ \nonumber
     && + 6 \left( \frac{s-u}{t-m^2}\right)^2
                          \bigl(\frac{m^2}{t}\bigr)^2
                                          \Biggr] \ .
\end{eqnarray}

As in the massless case, the cross section for $\bar q +g \to \bar q + G_{KK}$ is also
 the same as that of $q +g \to q + G_{KK}$. 
  For the process
 $g+g \to g+ G_{KK}$, one finds 
 \begin{eqnarray}
\frac{d\sigma}{d \cos\theta}(g+g \to g+ G_{KK})&=& \frac{3}{16}
      \frac{\pi \alpha_s G_N}{(1-m^2/s)(1-\cos^2\theta)} \times
            \\ \nonumber
     && 
      \Biggl[(3 + 
   \cos^2\theta )^2\left(1 + \bigl(\frac{m^2}{s}\bigr)^4\right) - 4 ( 7 + \cos^4\theta)\frac{m^2}{s }\left(1 +  
           \bigl(\frac{m^2}{s}\bigr)^2\right)  \\ \nonumber
     && 
           + 6 (9 - 2  \cos^2\theta +  \cos^4\theta)  
                      \bigl(\frac{m^2}{s}\bigr)^2  
                                          \Biggr] \ .
\end{eqnarray}
It is easy to see that one recovers the massless graviton case by taking the mass of the KK graviton to zero. Note that in the ADD case, to obtain the collider cross sections, one has to  also perform an integration over the KK-mass distributions as well as the usual integration over the 
parton densities. Of course in the case of the four dimensional model no KK sum is required. 
To be specific we will employ the CTEQ6.6 parton distributions{\cite {cteq}} in obtaining the results obtained below.  

\begin{figure}
\centering
\includegraphics[width=11cm,angle=90]{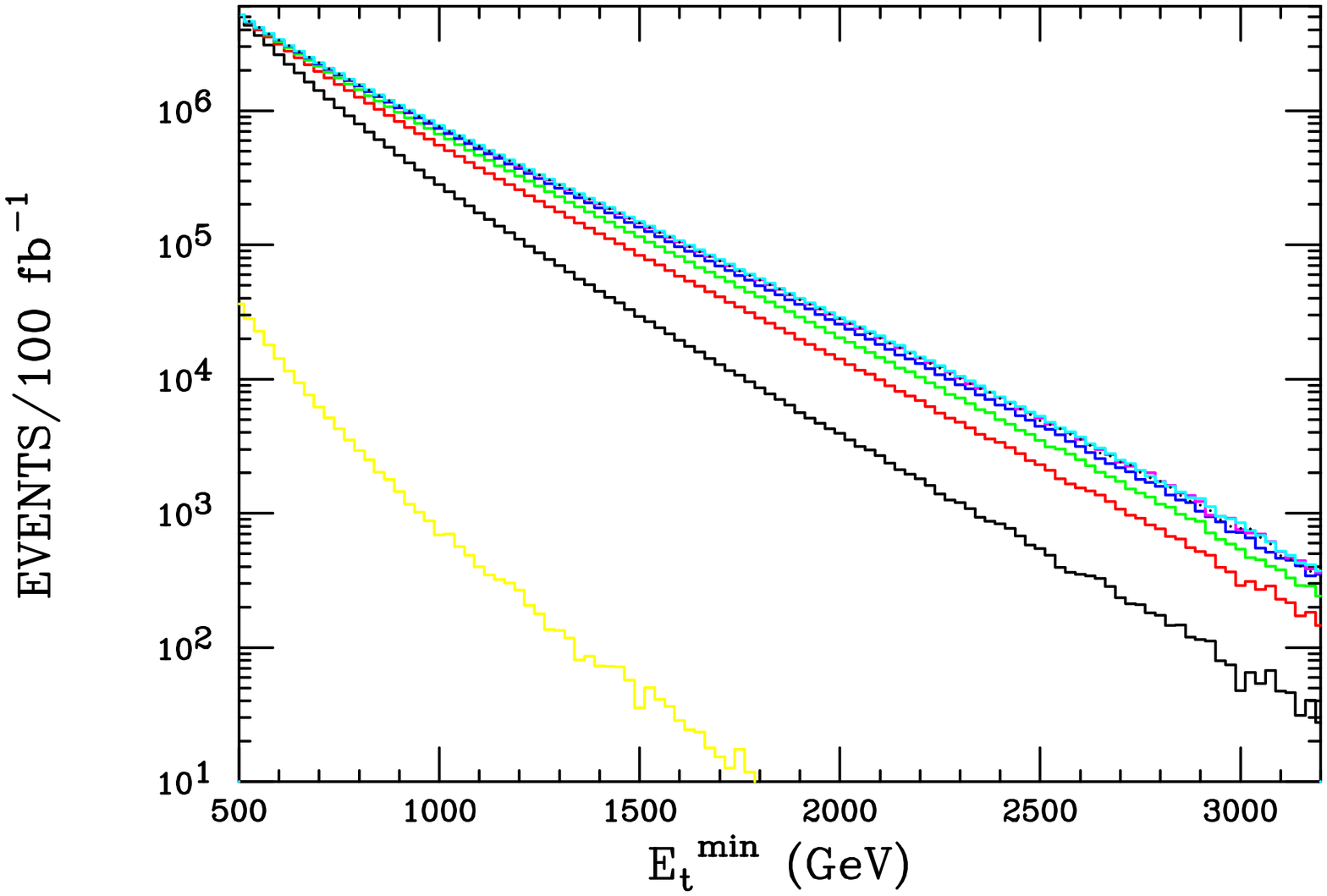}
\caption{This figure shows the shape of the $E_T^{min}$ distribution counting the number of monojet events at the 14 TeV LHC assuming a luminosity 
of 100 fb$^{-1}$ 
above a $E_T^{min}$ cut of 500 GeV and requiring a central jet $|\eta_j| <3$. The yellow histogram is the expected standard model 
background as discussed in the text while the red and higher histograms are for the ADD model with the number of extra dimensions being 2,3,4, etc. 
The lower solid black histogram is for the four dimensional model with $M_P$=1 TeV. The ADD results were in each case adjusted by varying their 
associated Planck scale to produce the same result as does the four dimensional model at $E_T^{min}=500$ GeV in order to show the relative shapes for 
these distributions.}
\label{Figure1}
\end{figure}

Using the partonic cross sections above we can employ a modified version of the code developed for \cite{Rizzo:2008vr}, 
and can then generate events 
for the 14 TeV LHC for both the ADD and the four dimensional models as well as for the 
the standard model background. Here, following Ref.{\cite {hinch}}, we expect this background 
to be dominated by the production of $Z$ plus a single jet with the $Z$ decaying into pairs of neutrinos. 
This background can be much reduced by requiring a missing energy and/or jet energy cut of at least 500 GeV 
and demanding that the jet to be central $|\eta_j|<3$. The results of our direct comparison of the ADD predictions with those of the four 
dimensional model assuming, \eg, that $M_P=1$ TeV can be found in Fig. \ref{Figure1}. Here we see that the two 
new physics models predict monojet $E_T$ distributions which are reasonably dissimilar in overall shape. The falling 
four dimensional model monojet spectrum above the $\sqrt {\hat s}=M_P$ threshold is seen to be somewhat softer 
than the corresponding ADD model prediction for any number of extra dimensions. 

\begin{figure}[htp]
\centering
\includegraphics[width=11cm,angle=90]{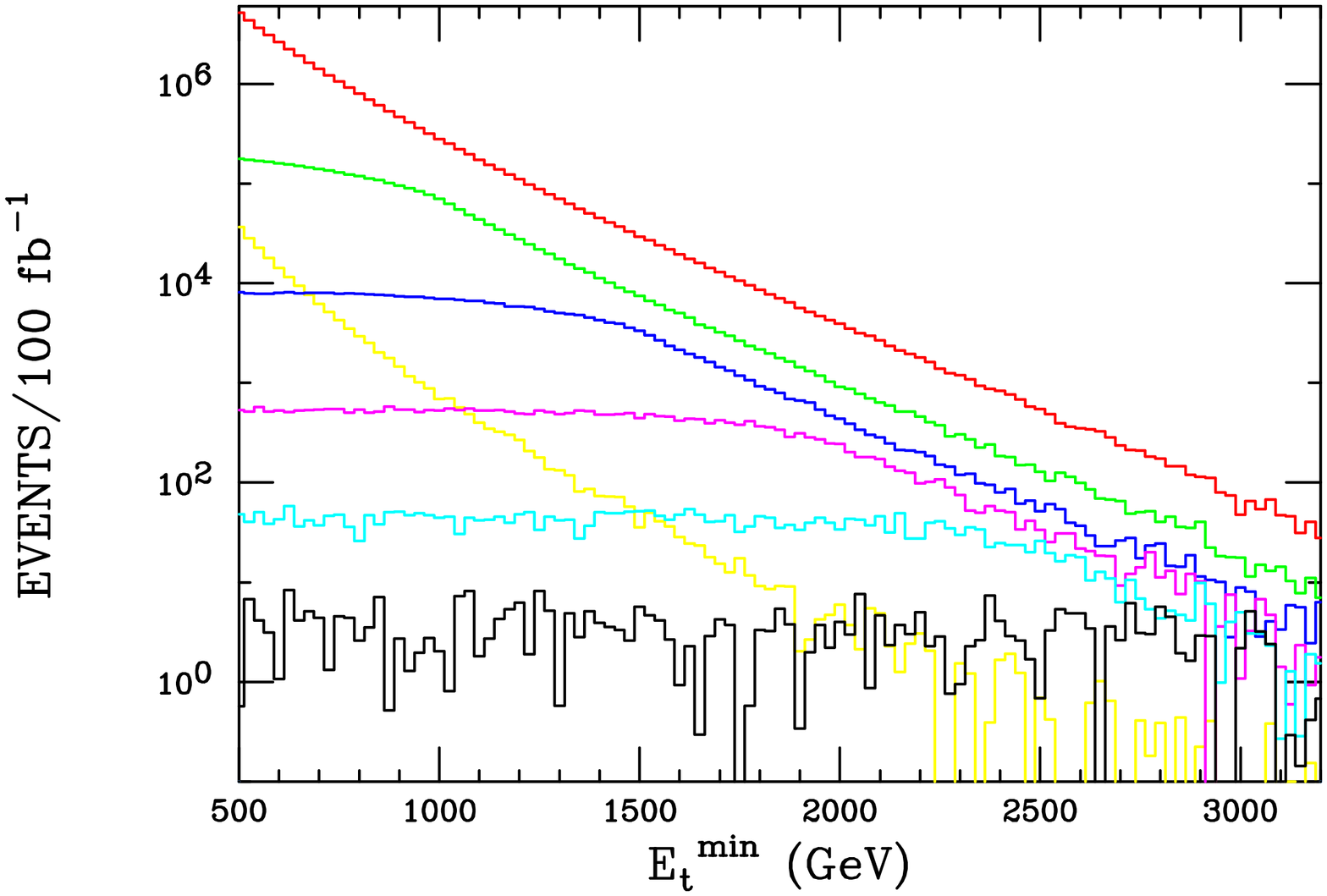}
\caption{This figure for the 14 TeV LHC shows the event rate for the standard model jet+missing energy background as a function of the cut on the jet $E_T$ 
in yellow as well as the four dimensional model predictions for the cases $M_{P}$=1(2,3,4,5,6) TeV from top to bottom in red, green, blue,...  
and requiring that the existence of a threshold at $\sqrt {\hat s}= M_P$. From this figure, one can deduce that 
the search reach for the four dimensional model at the LHC is $\simeq$ 5 TeV for a luminosity =100 fb$^{-1}$. The 
shape of the signal histograms with the requirement above are quite different than those for ADD due to the vanishing of the cross 
section at small $\hat s$. Note the shape change at $E_T^{min}=0.5 M_P$ which is a result of this cross section threshold that is  
absent in the ADD model.}
\label{Figure2}
\end{figure}

\begin{figure}
\centering
\includegraphics[width=11cm,angle=90]{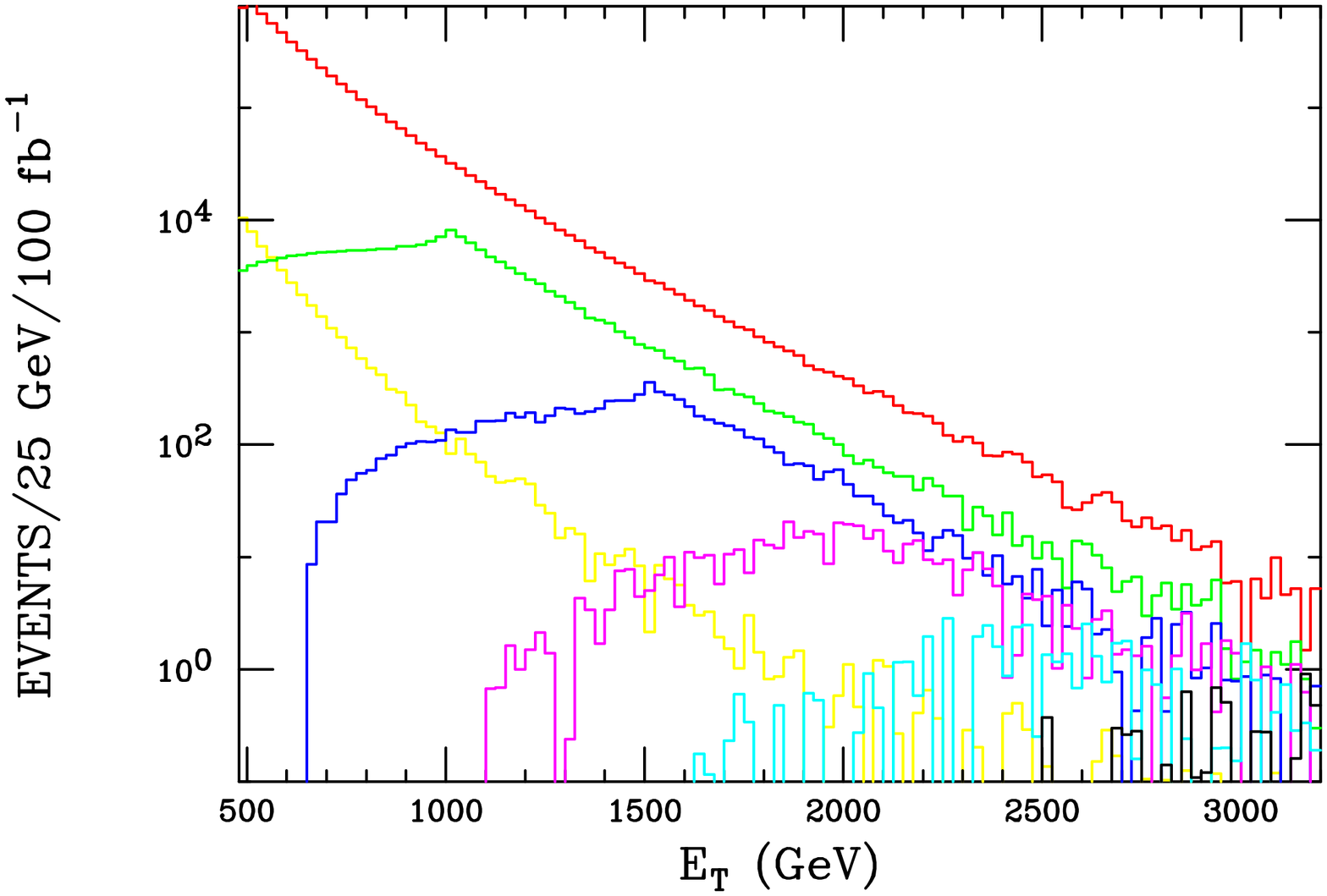}
\caption{This figure shows the monojet $E_T$ distributions at the 14 TeV the LHC assuming a luminosity of 100 fb$^{-1}$. The histograms are 
as labeled as in the previous figure. Note the kink-like structure in the four dimensional model distributions at $E_T= 0.5 M_P$ which is a result of 
the cross section threshold that is absent in the ADD model.}
\label{Figure3}
\end{figure}

From Figs.\ref{Figure2} and \ref{Figure3} we can obtain an estimate of the search reach for the four dimensional model at the 
14 TeV LHC of $\simeq 5$ TeV assuming an integrated luminosity of 100 $fb^{-1}$. Here we also see some unusual features in these 
distributions associated with the four dimensional model which are absent from the case of ADD which are due to the cross section 
threshold at $\sqrt {\hat s}=M_P$. Unlike for the ADD case, whose $E_T$ and $E_T^{cut}$ distributions fall monotonically, the 
threshold in the four dimensional model cross section naturally leads to structure in these corresponding distributions. This 
added kink-like structure is a clear aid in distinguishing the predictions of these two classes of models at the LHC as can easily be seen 
from these figures. For the case of the $E_T^{cut}$ distribution we see that it is essentially flat until the value $E_T^{cut}=0.5M_P$ 
is reached and then falls monotonically. On the otherhand, the $E_T$ distribution rises below the value of $E_T=0.5M_P$ at which point a peak occurs. 
At higher values of $E_T$ the distribution fall monotonically. These are quite distinctive indications of a cutoff in the cross section 
at a fixed value of $\sqrt {\hat s}$ and would allow us to extract this value directly from the LHC experimental data.  

Here we have focussed on one aspect of quantum gravity phenomenology at the LHC. Other interesting signatures are small semi-classical black holes \cite{Dimopoulos:2001hw} or small non-thermal black holes \cite{Calmet:2008dg}. Although small black holes would be easier to observe than the missing energy signals discussed above, the theoretical framework typically used to discuss the formation of small holes at the LHC is less reliable than the one used here for the emission of  gravitons which relies on the well understood expansion of the Hilbert-Einstein action around the Minkowski metric. This approach thus assumes very little about the nature of quantum gravity and the bounds derived that way will therefore be much more reliable than using small black holes.

The LHC will start to operate in the coming months. Although it may take time to reach its design energy and luminosity, it is quite remarkable to realize  that once does it will be able to probe not only whether the solution to the hierarchy problem is based on low scale quantum gravity, but it will  also give us the possibility to learn whether a low Planck mass is due to large extra-dimensions or a large hidden sector with {\it only} four space-time dimensions. The results presented in this work will be useful even if the solution to the gauge hierarchy problem is not based on gravity. In that case, the LHC will be able to set the tightest limit to date on the value of the energy scale at which quantum gravitational effects become strong and this in a model independent way.

\bigskip

{\it Acknowledgments:} 
The work of TGR was supported in part by the Department of Energy, Contract DE-AC02-76SF00515. 
%\newpage

%%%%%%%%%%%%%%%%%%%%%%%%%%%%%%%%%%%%%%%%%%%%%%%%%%%%%%%%%%%%%%%%%
%%%
%%%                     BIBLIOGRAPHY
%%%
%%%%%%%%%%%%%%%%%%%%%%%%%%%%%%%%%%%%%%%%%%%%%%%%%%%%%%%%%%%%%%%%%

\bigskip

%\newpage
%\vskip .75 in


\baselineskip=1.6pt
\begin{thebibliography}{99}
%\cite{ArkaniHamed:1998rs}

\bibitem{ArkaniHamed:1998rs}
  N.~Arkani-Hamed, S.~Dimopoulos and G.~R.~Dvali,
  %``The hierarchy problem and new dimensions at a millimeter,''
  Phys.\ Lett.\  B {\bf 429}, 263 (1998)
  [arXiv:hep-ph/9803315];
  %%CITATION = PHLTA,B429,263;%%

%\cite{Randall:1999ee}
\bibitem{Randall:1999ee}
  L.~Randall and R.~Sundrum,
  %``A large mass hierarchy from a small extra dimension,''
  Phys.\ Rev.\ Lett.\  {\bf 83}, 3370 (1999)
  [arXiv:hep-ph/9905221].
  %%CITATION = PRLTA,83,3370;%%

      %\cite{Calmet:2008tn}
\bibitem{Calmet:2008tn}
X.~Calmet, S.~D.~H.~Hsu and D.~Reeb,
  %``Quantum gravity at a TeV and the renormalization of Newton's constant,''
  Phys.\ Rev.\  D {\bf 77}, 125015 (2008)
  [arXiv:0803.1836 [hep-th]].
  %%CITATION = PHRVA,D77,125015;%%

\bibitem{Dvali}
%\cite{Dvali:2001gx}
%\bibitem{Dvali:2001gx}
  G.~R.~Dvali, G.~Gabadadze, M.~Kolanovic and F.~Nitti,
  %``Scales of gravity,''
  Phys.\ Rev.\  D {\bf 65}, 024031 (2002)
  [arXiv:hep-th/0106058];
  %%CITATION = PHRVA,D65,024031;%%
%\cite{Dvali:2007hz}
%\bibitem{Dvali:2007hz}
  G.~Dvali,
  %``Black Holes and Large N Species Solution to the Hierarchy Problem,''
  arXiv:0706.2050 [hep-th];
  %%CITATION = ARXIV:0706.2050;%%
%\cite{Dvali:2007wp}
%\bibitem{Dvali:2007wp}
  G.~Dvali and M.~Redi,
  %``Black Hole Bound on the Number of Species and Quantum Gravity at LHC,''
  Phys.\ Rev.\  D {\bf 77}, 045027 (2008)
  [arXiv:0710.4344 [hep-th]].
  %%CITATION = PHRVA,D77,045027;%%

%\cite{Larsen:1995ax}
\bibitem{Larsen:1995ax}
  F.~Larsen and F.~Wilczek,
  %``Renormalization of black hole entropy and of the gravitational coupling
  %constant,''
  Nucl.\ Phys.\  B {\bf 458}, 249 (1996)
  [arXiv:hep-th/9506066].
  %%CITATION = NUPHA,B458,249;%%

   %\cite{Rizzo:2008vr}
\bibitem{Rizzo:2008vr}
  T.~G.~Rizzo,
  %``Unique Signatures of Unparticle Resonances at the LHC,''
  JHEP {\bf 0811}, 039 (2008)
  [arXiv:0809.4659 [hep-ph]].

\bibitem{georgi}
%\cite{Georgi:2007si}
%\bibitem{Georgi:2007si}
  H.~Georgi,
  %``Another Odd Thing About Unparticle Physics,''
  Phys.\ Lett.\  B {\bf 650}, 275 (2007)
  [arXiv:0704.2457 [hep-ph]] and 
  %%CITATION = PHLTA,B650,275;%%
%\cite{Georgi:2007ek}
%\bibitem{Georgi:2007ek}
%  H.~Georgi,
  %``Unparticle Physics,''
  Phys.\ Rev.\ Lett.\  {\bf 98}, 221601 (2007)
  [arXiv:hep-ph/0703260].
  %%CITATION = PRLTA,98,221601;%%

%\cite{Cheung:2008xu}
\bibitem{Cheung:2008xu}
  K.~Cheung, W.~Y.~Keung and T.~C.~Yuan,
  %``Unparticle Phenomenology -- A Mini Review,''
  AIP Conf.\ Proc.\  {\bf 1078}, 156 (2009)
  [arXiv:0809.0995 [hep-ph]].
  %%CITATION = APCPC,1078,156;%%

  %\cite{Calmet:2009gn}
\bibitem{Calmet:2009gn}
  X.~Calmet and P.~de Aquino,
  %``Quantum Gravity at the LHC,''
  arXiv:0906.0363 [hep-ph].
  %%CITATION = ARXIV:0906.0363;%%

%\cite{Mirabelli:1998rt}
\bibitem{Mirabelli:1998rt}
  E.~A.~Mirabelli, M.~Perelstein and M.~E.~Peskin,
  %``Collider signatures of new large space dimensions,''
  Phys.\ Rev.\ Lett.\  {\bf 82}, 2236 (1999)
  [arXiv:hep-ph/9811337]; 
  %%CITATION = PRLTA,82,2236;%%

%\cite{Giudice:1998ck}
%\bibitem{Giudice:1998ck}
  G.~F.~Giudice, R.~Rattazzi and J.~D.~Wells,
  %``Quantum gravity and extra dimensions at high-energy colliders,''
  Nucl.\ Phys.\  B {\bf 544}, 3 (1999)
  [arXiv:hep-ph/9811291].
  %%CITATION = NUPHA,B544,3;%%

\bibitem{cteq}
%\cite{Nadolsky:2008zw}
%\bibitem{Nadolsky:2008zw}
  P.~M.~Nadolsky {\it et al.},
  %``Implications of CTEQ global analysis for collider observables,''
  arXiv:0802.0007 [hep-ph].
  %%CITATION = ARXIV:0802.0007;%%

\bibitem{hinch}
%\cite{Vacavant:2001sd}
%\bibitem{Vacavant:2001sd}
  L.~Vacavant and I.~Hinchliffe,
  %``Signals Of Models With Large Extra Dimensions In Atlas,''
  J.\ Phys.\ G {\bf 27}, 1839 (2001).
  %%CITATION = JPHGB,G27,1839;%%

%\cite{Dimopoulos:2001hw}
\bibitem{Dimopoulos:2001hw}
  S.~Dimopoulos and G.~L.~Landsberg,
  %``Black Holes at the LHC,''
  Phys.\ Rev.\ Lett.\  {\bf 87}, 161602 (2001)
  [arXiv:hep-ph/0106295].
  %%CITATION = PRLTA,87,161602;%%
  
  %\cite{Calmet:2008dg}
\bibitem{Calmet:2008dg}
  X.~Calmet, W.~Gong and S.~D.~H.~Hsu,
  %``Colorful quantum black holes at the LHC,''
  Phys.\ Lett.\  B {\bf 668}, 20 (2008)
  [arXiv:0806.4605 [hep-ph]].
  %%CITATION = PHLTA,B668,20;%%

  
\end{thebibliography}
\end{document}